\documentclass[reprint,amssymb, amsmath, aps, superscriptaddress, showpacs, footinbib, prl]{revtex4-1}
\usepackage[T1]{fontenc}
\usepackage[utf8]{inputenc}
\usepackage{amssymb}
\usepackage{amsmath}
\usepackage{color}
\usepackage{graphicx}
\usepackage{footmisc,booktabs}
\usepackage{gensymb}
\usepackage{sidecap}

\begin{document}

\title{Emergent Weyl fermion excitations in TaP explored by $^{181}$Ta quadrupole resonance}

\author{H. Yasuoka}
\affiliation{Max Planck Institute for Chemical Physics of Solids, 01187 Dresden, Germany}
\affiliation{Advanced Science Research Center, Japan Atomic Energy Agency, Tokai, Ibaraki 319-1195, Japan}

\author{T. Kubo}
\affiliation{Max Planck Institute for Chemical Physics of Solids, 01187 Dresden, Germany}
\affiliation{Department of Physics, Graduate school of Science, Kobe University, 1-1 Rokkodai, Nada-ku, Kobe, 657-8501 Japan}

\author{Y. Kishimoto}
\affiliation{Max Planck Institute for Chemical Physics of Solids, 01187 Dresden, Germany}
\affiliation{Department of Physics, Graduate school of Science, Kobe University, 1-1 Rokkodai, Nada-ku, Kobe, 657-8501 Japan}

\author{D. Kasinathan}
\affiliation{Max Planck Institute for Chemical Physics of Solids, 01187 Dresden, Germany}

\author{M. Schmidt}
\affiliation{Max Planck Institute for Chemical Physics of Solids, 01187 Dresden, Germany}

\author{B. Yan}
\affiliation{Max Planck Institute for Chemical Physics of Solids, 01187 Dresden, Germany}

\author{Y. Zhang}
\affiliation{Max Planck Institute for Chemical Physics of Solids, 01187 Dresden, Germany}
\affiliation{IFW Dresden, P.O. Box 270116, 01171 Dresden, Germany}

\author{H. Tou}
\affiliation{Department of Physics, Graduate school of Science, Kobe University, 1-1 Rokkodai, Nada-ku, Kobe, 657-8501 Japan}

\author{C. Felser}
\affiliation{Max Planck Institute for Chemical Physics of Solids, 01187 Dresden, Germany}

\author{A. P. Mackenzie}
\affiliation{Max Planck Institute for Chemical Physics of Solids, 01187 Dresden, Germany}
\affiliation{SUPA, School of Physics \& Astronomy, University of St. Andrews,  North Haugh, St. Andrews KY16 9SS, United Kingdom }

\author{M. Baenitz}
\email{Michael.Baenitz@cpfs.mpg.de}
\affiliation{Max Planck Institute for Chemical Physics of Solids, 01187 Dresden, Germany}

\date{\today}

\begin{abstract}
The $^{181}$Ta quadrupole resonance (NQR) technique has been utilized to investigate the microscopic 
magnetic properties of the Weyl semi-metal TaP. We found three zero-field NQR signals associated with 
the transition between the quadrupole split levels for Ta with $I$=7/2 nuclear spin. A quadrupole 
coupling constant, $\nu_{\mathrm{Q}}$ =19.250 MHz, and an asymmetric parameter of the electric 
field gradient, $\eta$ = 0.423 were extracted, in good agreement with band structure calculations. 
In order to examine the magnetic excitations, the temperature dependence of the spin lattice relaxation 
rate (1/$T_{\mathrm{1}}T$) has been measured for the $f_{\mathrm{2}}$-line ($\pm$5/2 $\leftrightarrow$
 $\pm$3/2 transition). We found that there exists two regimes with quite different relaxation processes. 
Above $T\text{*}$ $\approx$ 30\,K, a pronounced (1/$T_{\mathrm{1}}T$) $\propto$ $T^{2}$ behavior was found,
which is attributed to the magnetic excitations at the Weyl nodes with temperature dependent orbital 
hyperfine coupling. Below $T\text{*}$, the relaxation is mainly governed by Korringa process with
1/$T_{\mathrm{1}}T$ = constant, accompanied by an additional $T^{-1/2}$ type dependence to fit our experimental data.
We show that Ta-NQR is a novel probe for the bulk Weyl fermions and their excitations.

\end{abstract}


\maketitle

The past decade has seen an explosion of interest in the role of topology in condensed matter physics. Major discoveries 
have included the two dimensional graphene\cite{Castro_2009} and the topological insulators (TI) (e.g. HgTe or Bi$_2$Se$_3$),\cite{Kane_2005,Bernevig_2006,Hasan_2010}
 whose topological properties require the existence of gapless surface states. Many of the new materials host exotic excitations 
 whose observation can be regarded as direct experimental evidence for the existence of quasiparticles.
Arguably, the most topical of the new classes of materials are Dirac- and Weyl-semi metals which are predicted to host topologically 
 protected states in the bulk.\cite{Wehling_2014} In Dirac semimetals (DSM),\cite{Wehling_2014,Young_2012,Wang_2013} (e.g. Cd$_2$As$_3$ or Na$_3$Bi) each node contains fermions of two 
 opposite chiralities, whereas in the Weyl semimetals (WSM),\cite{Weng_2015,Shekar_2015,Sun_2015,Xu_2015,Arnold_2016} an even more interesting situation arises. A combination of non-centrosymmetric 
 crystal structure and sizable spin-orbit coupling (SOC) causes the nodes to split into pairs of opposite chirality (Weyl points). In the ideal case, 
 there would be exactly half filling of the relevant bands, such that the Weyl points would sit at the Fermi level ($E_{\mathrm{F}}$) and the 
 Weyl fermions would be massless. In actuality, Weyl semimetals such as the $d$-electron monophosphides NbP and TaP, 
 $E_{\mathrm{F}}$ does not exactly coincide with the Weyl nodes.\cite{Shekar_2015,Sun_2015,Arnold_2016} However, if the
nodes sit close enough to $E_{\mathrm{F}}$, in a region of linear dispersion ($E$ $\propto$ $k$), the Weyl physics can still
be observed in the excitations in the energy window $k_{\mathrm{B}}T$. A key issue in the study of the monophosphides is therefore to establish 
how close to the Fermi level the Weyl points sit, and to estimate the range of energy over which the linear dispersion exists. This presents 
a considerable experimental challenge. The nodes appear in the electronic structure of the bulk, and the materials are fully three-dimensional, 
so the surface-sensitive techniques that have yielded immense insight into other topological physics are not ideally suited to studying the Weyl points. 
Primarily, one would like to identify a bulk probe that can excite the Weyl  fermions and probe the linear dispersion $E$ $\propto$ $k$
indirectly via its energy dependence of the density of states around the Fermi level, which is $N(E) \propto E^{\mathrm{2}}$ for a Weyl node.\cite{Wehling_2014}
The magnetic resonance method in general has the ability to probe $N(E)$, and was applied successfully to systems 
like unconventional superconductors (e.g. UPt$_3$)\cite{Curro_2011,Curro_2009,Walstedt_2008,Kuramoto_2000} or correlated magnetic semimetals\cite{Aeppli_1992,*Riseborough_2000} (e.g. SmB$_6$\cite{Caldwell_2007}  or CeRu$_4$Sn$_6$\cite{Bruning_2010}). 
In particular, for unconventional superconductors, the nuclear quadrupole resonance (NQR) spin lattice relaxation provides information 
about $N(E)$ around the $E_{\mathrm{F}}$
 and allows us to distinguish between point nodes ($N(E) \propto E^{\mathrm{2}}$) and line nodes ($N(E) \propto E$).\cite{Kuramoto_2000}
 Therefore,
 NQR should be a good tool to study the low energy spin excitations in a Weyl semi metal and is the focus of our presented work. Assuming that some of the 
 Weyl points are energetically not too far from $E_{\mathrm{F}}$, NQR can probe the magnetic excitations of emergent Weyl fermions via the temperature dependence of the spin-lattice relaxation rate (1/T$_{1}T$). Furthermore, a characteristic temperature dependence of the hyperfine coupling between the nuclear spin and electric orbitals near the Weyl nodes has theoretically been predicted which modifies the temperature dependence of (1/T$_{1}T$) in a special manner.\cite{Okvatovity_2016}
 In fact, TaP has been known to have two sets of Weyl nodes, one located 41 meV (476\,K) below $E_{\mathrm{F}}$  and the other is located 13 meV (151\,K) above 
 $E_{\mathrm{F}}$.\cite{Arnold_2016} Accordingly, for temperatures coinciding with the Weyl nodes, excitations associated with the Weyl fermions are expected.   Here, we present such characteristic excitations via Ta NQR experiments and explore the emergent Weyl fermion excitations in TaP.

\begin{figure}[t]
\includegraphics[clip,width=\columnwidth]{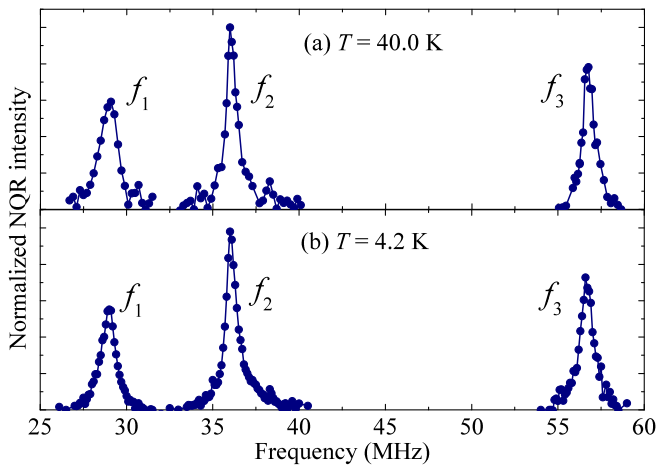}
\caption{Typical $^{181}$Ta NQR spectra in TaP obtained from the spin-echo real part integration at 40\,K (a) and 4.2\,K (b). The data was taken
in 0.1\,MHz steps across the spectrum. The lines $f_{\mathrm{1}}$, $f_{\mathrm{2}}$ and $f_{\mathrm{3}}$ correspond to 
$\pm$3/2 $\leftrightarrow$ $\pm$1/2, $\pm$5/2 $\leftrightarrow$ $\pm$3/2 and $\pm$7/2 $\leftrightarrow$ $\pm$5/2 transitions, respectively.
}
\end{figure}

Samples used in the present NQR study were prepared by the chemical transport method. In a first step, TaP was synthesized by 
direct reaction of the elements tantalum (Alfa Aesar 99.98\%) and red phosphorus (Alfa Aesar 99.999\%) at 500 $\degree$C and 600 $\degree$C 
in an evacuated fused silica tube for 72 hours. Starting from this microcrystalline powder, TaP was crystallized by a chemical transport reaction (CTR) 
in a temperature gradient from 900 $\degree$C (source) to 1000 $\degree$C (sink), and a transport agent concentration of 13 mg/cm$^3$ iodine 
(Alfa Aesar 99.998\%).  Crystals obtained by the CTR method were characterized by electron-probe-microanalysis and powder X-ray diffraction.

The NQR experiments were carried out with either a high quality single crystal or powder prepared from single crystals.
The NQR spectra and 1/$T_{1}T$ were measured using standard pulsed NMR (nuclear magnetic resonance) apparatus. The spectra were taken using the frequency sweep 
method under zero applied magnetic field. In order to avoid any artificial broadening, fast Fourier transformed (FFT) signals were summed across the 
spectrum (FFT-summation) or the real part was integrated after proper phase adjustment. Since $T_1$ is extremely long in TaP (typically several hundred 
seconds at low temperatures), we employed the progressive saturation method to measure the temperature dependence of 1/$T_{1}T$.\cite{Mit_2001} 
The recovery of nuclear magnetization was fitted to the theoretical function\cite{Chepin_1991} for the magnetic relaxation in NQR lines,
\begin{eqnarray}
M_{n}(t) &=& M_{0} \left[1-\left(Q_{1} e^{(-K_{1}t/T_{1})} + Q_{2} e^{(-K_{2}t/T_{1})} \right. \right. \nonumber \\
               &+& \left. \left. Q_{3} e^{(-K_{3}t/T_{1})}\right)\right]
\end{eqnarray}
where, $Q_n$ and $K_n$ are constants depending on which NQR transition is excited and the asymmetry parameter of the 
electric field gradient (EFG). Using a set of principal axes, the quadrupole Hamiltonian can be written as,\cite{Abragam_1961,*Slichter_1989} 
\begin{equation}
H_{Q} = \frac{e^{2}qQ}{4I(2I-1)} \left[ 3I_{z}^{2} - I(I+1) + \frac{1}{2}\eta (I_{+}^{2} - I_{-}^{2}) \right]
\end{equation}
where, $eq$ is the largest component of the EFG tensor, $V_{\mathrm{zz}}$, and $eQ$ is the nuclear quadrupole moment. 
The EFG tensor is generally defined as |$V_{\mathrm{zz}}$| $\geq$ |$V_{\mathrm{yy}}$| $\geq$ |$V_{\mathrm{xx}}$|, with the 
asymmetry parameter $\eta$ = ($V_{\mathrm{xx}}-V_{\mathrm{yy}}$)/$V_{\mathrm{zz}}$. The quadrupole-split nuclear energy levels $E_{m}$, and the resultant transition frequencies can be 
readily calculated numerically by diagonalizing Eq. (2). For $\eta$ = 0, the energy levels can simply be expressed as,
\begin{equation}
E_{m} = \frac{1}{6}h\nu_{Q} \left[ 3m^{2} - I(I+1)\right], \nu_{Q} = \frac{3e^{2}qQ}{h2I(I-1)} 
\end{equation}
where $\nu_{\mathrm{Q}}$ is the quadrupole coupling constant. The NQR occurs for the transition between two levels 
$m$ and $m$+1 and the resonance condition can be written as
$f_{\mathrm{Q}}$ = $\nu_{\mathrm{Q}}$(2|$m$|+1)/2.
 Therefore, for $\eta$ = 0, three NQR lines for $I$=7/2 are expected at $\nu_{\mathrm{Q}}$, 2$\nu_{\mathrm{Q}}$ and 3$\nu_{\mathrm{Q}}$ with equal spacing. 
 
\begin{figure}[t]
\includegraphics[clip,width=0.9\columnwidth]{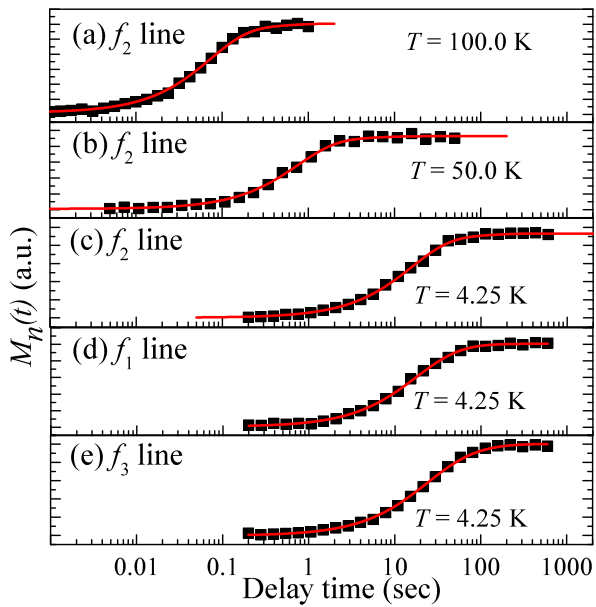}
\caption{(Color Online) Recovery of the nuclear magnetization $M_{\mathrm{n}}$(t) measured by the NQR intensity after saturation to
thermal equilibrium value for $f_{\mathrm{2}}$ line at 100\,K (a), 50\,K (b), 4.25\,K (c), $f_{\mathrm{1}}$ line at 4.25\,K (d) and
$f_{\mathrm{3}}$ line at 4.25\,K (e). The red curves are the least squares fit of the
data derived using Eq. (1). The perfect fit in all cases demonstrates the fact that the relaxation is governed by magnetic fluctuations. 
}
\end{figure}

We searched for the NQR signal in TaP powder in a frequency range from 20 MHz to 80 MHz at 4.2\,K and found three resonance lines. 
The observed $^{181}$Ta NQR spectra at 4.2\,K and 40\,K are shown in Fig.1 (a) and (b). One can immediately observe that the three lines are 
not equally spaced, meaning that the value of $\eta$ is sizable. By fitting each line at 4.2 K to a Lorentzian, we obtained the peak 
frequencies $f_{\mathrm{1}}$ = 28.95 MHz, $f_{\mathrm{2}}$ = 36.08 MHz and $f_{\mathrm{3}}$ = 56.67 MHz. Although the full width at 
half maximum (FWHM) of the spectra is about 800 KHz, the line profiles are Lorentzian with long spectral tails. This implies that the 
EFG has a rather broad distribution. Since the Ta nuclear quadrupole moment is quite large (besides rare-earth and 
actinide elements, the $^{181}$Ta nucleus has the second largest $Q$-value), the NQR spectrum can be broadened easily 
by merely a slight local inhomogeneity or nonstoichiometry in the composition. A least squares fit of the observed peak frequency to the theoretical quadrupole interaction 
obtained from exact diagonalization gives us $\nu_{\mathrm{Q}}$ = 19.250 MHz and $\eta$ = 0.423. The same results were also obtained 
for a small single crystal at 4.2 K. We also measured the spectra up to 80 K and found that there is no appreciable temperature dependence of 
the $\eta$ value but $\nu_{\mathrm{Q}}$ has a gradual decrease with increasing temperature (see Supplement for more details).

\begin{table}[t]
  \begin{tabular}{|c|c|c|c|}
  \hline
  \multicolumn{4}{|c|}{$K_{\mathrm{1}}$ = 3.01028, $K_{\mathrm{2}}$ = 8.7084, $K_{\mathrm{3}}$ = 17.42483} \\
  \hline
  $f_{\mathrm{1}}$-line & $Q_{\mathrm{1}}$ = 0.05073 & $Q_{\mathrm{2}}$ = 0.45677 & $Q_{\mathrm{3}}$ = 0.49250 \\
  \hline
   $f_{\mathrm{2}}$-line & $Q_{\mathrm{1}}$ = 0.07624 & $Q_{\mathrm{2}}$ = 0.02126 & $Q_{\mathrm{3}}$ = 0.90250 \\
   \hline
   $f_{\mathrm{3}}$-line & $Q_{\mathrm{1}}$ = 0.19320 & $Q_{\mathrm{2}}$ = 0.51185 & $Q_{\mathrm{3}}$ = 0.29495 \\
   \hline
   \end{tabular}
   \caption{The calculated pre-factors and exponents in Eq. (1) for $\eta$ = 0.423, used for the least squares fit shown in Fig.\,2.}
\end{table}  

In order to extract the quadrupole interaction in TaP theoretically, we performed band structure calculations using the 
density functional theory (DFT) code FPLO.\cite{Koepernik_1999} We used the Perdew-Wang parametrization of the local 
density approximation (LDA) for the exchange-correlation functional.\cite{Perdew_1992} The strong SOC in TaP is taken into account by performing 
full-relativistic calculations, wherein the Dirac Hamiltonian with a general potential is solved. The treatment of a finite nucleus
is implemented in the code, necessary for accurate estimation of NQR parameters.\cite{Koch_2010} As a basis set, we chose 
Ta (4$f$/5$s$5$p$6$s$7$s$8$s$5$d$6$d$7$d$6$p$7$p$5$f$) and P (2$s$2$p$3$s$4$s$5$s$3$p$4$p$5$p$3$d$4$d$4$f$) 
semi-core/valence states. The higher lying states of the basis set are essentially important for the calculation of the EFG tensor 
with components $V_{\mathrm{ij}}$ = $\partial V/ \partial x_{\mathrm{i}} \partial x_{\mathrm{j}}$. 
The low lying states were treated fully relativistically as core states. A well-converged $k$ mesh in 1210 $k$-points was used in 
the irreducible part of the Brillouin zone. Theoretically, the quadrupole coupling, $\nu_{\mathrm{Q}}$ can be obtained by calculating the
electric field gradient (EFG) at the Ta nuclear site which is defined as the second partial derivative of the electrostatic potential v(r) 
at the position of the nucleus $V_{\mathrm{ij}}$ = ($\partial \mathrm{i} \partial \mathrm{j}$v(0) - $\Delta \delta_{\mathrm{ij}}$v(0)/3). 
Our calculations result in $V_{\mathrm{xx}}$ = -1.186$\times$10$^{21}$ V/m$^{2}$, $V_{\mathrm{yy}}$ = -2.354$\times$10$^{21}$ V/m$^{2}$ 
and $V_{\mathrm{zz}}$ = 3.540$\times$10$^{21}$ V/m$^{2}$, where the principal axis of the EFG is [100] for one Ta atom and [010] for the second 
Ta atom in the unit cell. Using these values and Eq. (2) we obtained $\nu_{\mathrm{Q}}$ = 20.057 MHz and $\eta$ = 0.33. 
These theoretical values are in good agreement with the experimental values, assuring that our line assignment to the quadrupole transitions is correct.

Before discussing the temperature dependence of 1/$T_{\mathrm{1}}T$, we have to make sure that the relaxation is 
governed by the magnetic fluctuations associated with the conduction electrons, although the density is quite small in semi-metals. 
To answer this, we have made careful measurements of the time dependence of the recovery of nuclear magnetization from saturation 
to thermal equilibrium for all temperatures and NQR lines. Then, assuming that magnetic fluctuations are responsible for the 
nuclear relaxation process, the relaxation curves were fitted to Eq. (1) using calculated prefactors and exponents (Table. I) for the observed value 
of $\eta$ = 0.423. The experimental results and fitted curves are shown in Fig.\,2. Here, we have a perfect match between the two
for all temperatures and NQR lines, providing very strong evidence that the relaxation 
process is totally governed by magnetic fluctuations and yielding assurance to the accuracy of the $T_{\mathrm{1}}$ values that were extracted.

The temperature dependence of 1/$T_{\mathrm{1}}T$ has been measured mainly for the $f_{\mathrm{2}}$-line and the obtained result 
is shown in Fig.\,3(a). One can immediately observe that there exists a characteristic temperature $T\text{*}$ $\approx$ 30\,K, where the 
relaxation process has a crossover from a high temperature $T^{\mathrm{2}}$ behavior (which is presumably associated with the excitations in 
the nodal structure of Weyl points) to the low temperature Korringa excitations\cite{Korringa_1950} for parabolic bands ($E \propto k^{\mathrm{2}}$ and 
$N(E) \propto \sqrt{E}$) with a weak temperature dependence. 

\begin{figure}[t]
\includegraphics[clip,height=6.2cm]{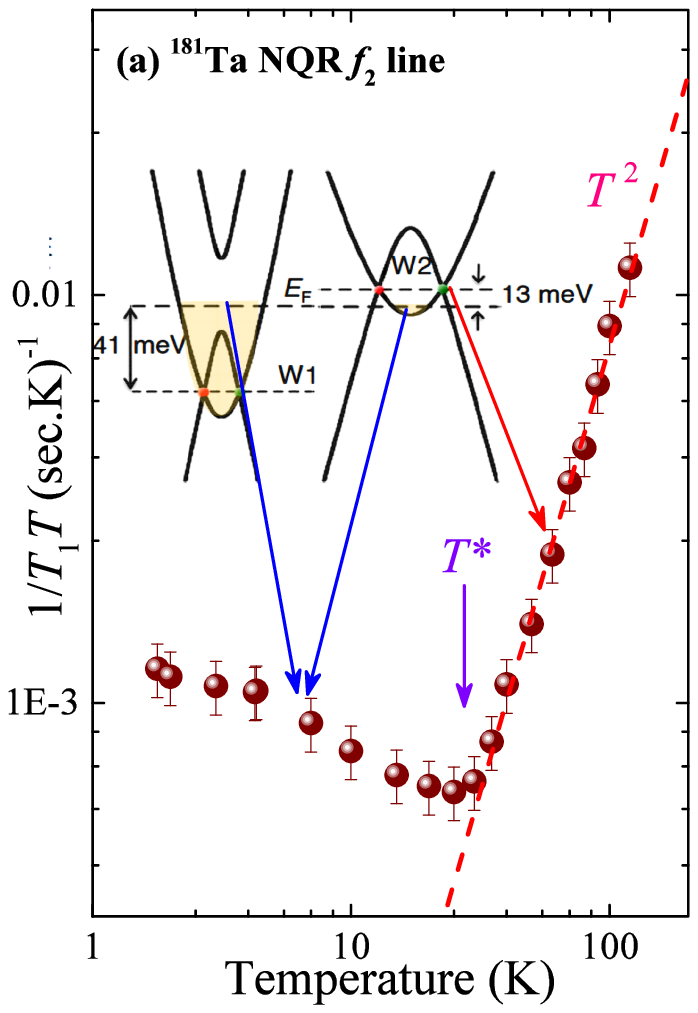}
\includegraphics[clip,height=6.2cm]{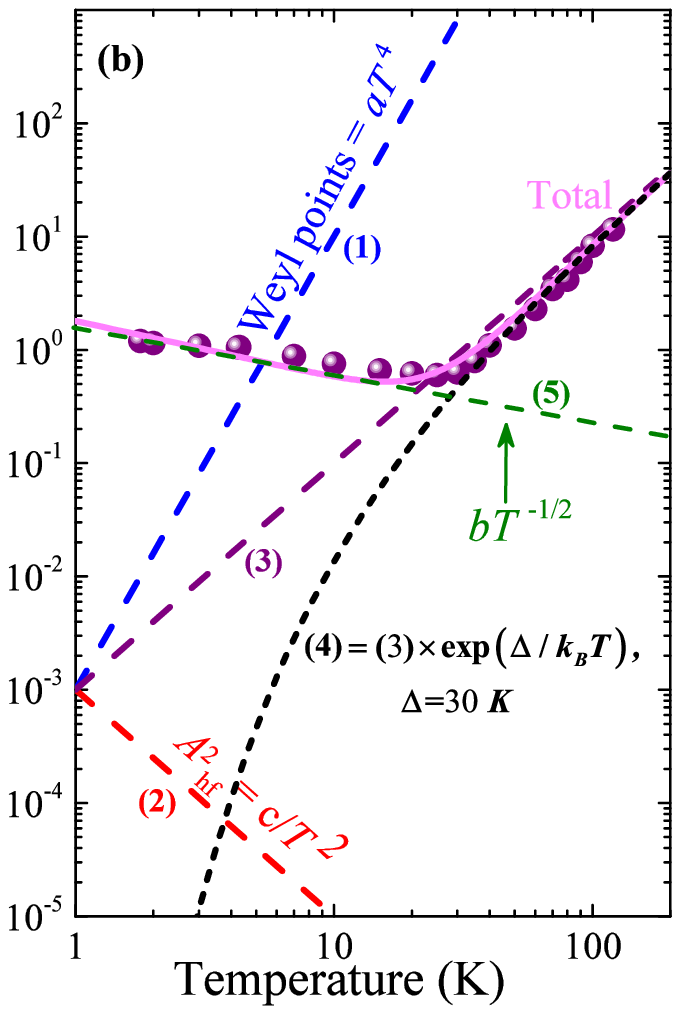}
\caption{(Color Online) (a) The temperature dependence of 1/$T_{\mathrm{1}}T$ of TaP along with the schematics of the band structure. At
$T\text{*}$ $\approx$ 30\,K, the relaxation process crosses over between two different regimes. (b) Schematic illustration of Eq. (5): Curves (1), (2) and (3) are
expected temperature dependence from the second term (Weyl nodes) not including the exponent, while curve (4) includes the exponent. The $T^{\mathrm{-1/2}}$
dependence associated with the conventional bands is depicted by curve (5). The magenta curve is the total relaxation behavior expected from Eq. (5) and
matches the experimental results. 
}
\end{figure}

Quite generally,  1/$T_{\mathrm{1}}T$ can be expressed using the wave vector ($q$) and frequency ($\omega$)
dependent magnetic susceptibility $\chi (q,\omega)$, characterizing the magnetic excitations in a system as,\cite{Katayama_2007}
\begin{equation}
\frac{1}{T_{1}T} = \frac{2\gamma_{n}^{2}k_{B}}{g^{2}\mu_{B}^{2}} \sum_q A_{q}^{2} \frac{\chi_{\perp}''(q,\omega)}{\omega_{n}}
\end{equation}
where $\chi_{\perp}''$($q,\omega$) is the transverse component of imaginary part of 
$\chi$($q,\omega$), $\gamma_{\mathrm{n}}$
is the nuclear gyromagnetic ratio and $A_{\mathrm{q}}$ is the $q$-dependent hyperfine coupling constant. 
Presently, a proper microscopic theory to accurately calculate $\chi$($q,\omega$) for multi-band systems is lacking. 
Consequently, we have evaluated a theoretical estimate for 1/$T_{\mathrm{1}}T$ based on non-interacting itinerant electrons approximation,
using standard DFT. 
In the band structure of TaP, besides the normal bands, two types of Weyl points appear. The first set of Weyl points, termed W1 and 
located in the $k_{\mathrm{z}}$ = 0 plane, and lie $\sim$\,40 meV below $E_{\mathrm{F}}$. 
The second set of Weyl points, W2, which lie nearly in the $k_{\mathrm{z}}$ = $\pi$/$c$ plane ($c$ is the lattice parameter along $z$) 
are $\sim$\,13 meV above $E_{\mathrm{F}}$.\cite{Arnold_2016} This feature is shown schematically in the insert of Fig.\,3(a) (see details in 
Supplement). Based on this band structure, one can easily imagine that the conventional Korringa process is valid 
for very low temperatures, with the upper bound limited by the energy of W2.
Upon further increasing the temperature, excitations at the Weyl node W2 should become progressively dominant. 
Then, 1/$T_{\mathrm{1}}T$ may be phenomenologically expressed in a two-channel relaxation model by
\begin{eqnarray}
\frac{1}{T_{1}T} &=& \frac{\pi k_{B}}{h}\left( A_{p}^{hf} \right)^{2}N_{p}(E)^{2} \nonumber \\
                             &+&  \alpha \left[ \left( A_{w}^{hf}\right)^{2}\int\left \langle\frac{N_{W}(E)^{2}}{N_{0}^{2}} \right \rangle f(E) \{1 - f(E)\} dE \right] \nonumber \\
                            &\cdot&  exp\left( \frac{-\Delta E}{k_{B}T} \right)
\end{eqnarray}
where, the first term corresponds to excitations associated with parabolic bands ($E \propto k^{\mathrm{2}}$) 
via the conventional Korringa process\cite{Korringa_1950} with the hyperfine coupling constant $A_{\mathrm{p}}^{\mathrm{hf}}$. The
second term is characteristic to the 
excitations of the Weyl nodes and linear bands ($E \propto k$) with $A_{\mathrm{w}}^{\mathrm{hf}}$. 
Herein, we have ignored the $q$ dependence. $\alpha$ is the scaling factor. 
The term 
$N_{\mathrm{w}}(E)$ = $N_{\mathrm{0}}E/[E^{2}-\Delta(\theta,\phi)$]$^{1/2}$ depends on the nodal structure and $f(E)$ is the Fermi distribution function. 
Because of the 
gap ($\Delta E$) between W2 and $E_{\mathrm{F}}$, we include an activation term, $exp(\Delta E/k_{\mathrm{B}}T)$ in the second process. 
In general, for the point nodal case, we know $\Delta(\theta, \phi) = \Delta_{0}sin\theta$, and $N(E) \propto E^{2}$; accordingly 1/$T_{\mathrm{1}}T$ $\propto$ $T^{4}$, which was observed experimentally for a point node superconductor.\cite{Katayama_2007}
Recently, anomalous hyperfine coupling due to orbital magnetism in the Weyl node has been predicted theoretically,\cite{Okvatovity_2016}
where the orbital contribution to $A_{\mathrm{w}}^{\mathrm{hf}}$ has a 1/$T$ dependence. Then, the second term of Eq.\,(5) becomes $T^{2}exp(-\Delta E/k_{\mathrm{B}}T)$. For the first term ($T$ < 30\,K), experimental data shows a $T^{-1/2}$ temperature dependence, despite the temperature independence of
the Korringa process. The origin of this is not clear but we may speculate that correlation among excited quasiparticles may affect it. 
Setting $\Delta E/k_{\mathrm{B}} \approx T\text{*}$ (30\,K), the expected temperature dependence is obtained by summing up the above contributions. 
As schematically shown in Fig 3. (b), choosing $\alpha$ = 10$^{-3}$, Eq. (5) 
is in fairly good agreement with the experiment. In particular, we clearly see a $T^{2}$ dependence for $T > T\text{*}$ which we believe to be the manifestation of 
Weyl fermion excitations near the Weyl points in TaP. 

In conclusion, we have reported the observation of a complete set of $^{181}$Ta NQR lines in Weyl semimetal TaP. All observed NQR lines are 
consistently assigned to the transitions between $m$- and ($m$+1)- states ($m$ = $\pm$5/2, $\pm$3/2 and $\pm$1/2). 
From our measurements, we obtain an asymmetry parameter $\eta$ = 0.423 and a quadrupole coupling constant of $\nu_{\mathrm{Q}}$ = 19.250 MHz. These
findings are in good agreement with DFT calculations which provide $\eta$ = 0.33 and $\nu_{\mathrm{Q}}$ = 20.057 MHz. 
The low energy excitations as a function of temperature were probed through the Ta spin lattice relaxation rate (1/$T_{\mathrm{1}}T$) 
which shows a pronounced $T^{\mathrm{2}}$ behavior above $T\text{*}$ = 30\,K and a $T^{-1/2}$ behavior below $T\text{*}$. 
The relaxation process below $T\text{*}$ is mostly related to the conventional density of states ($N(E) \propto \sqrt{E}$) which yields an almost constant
density of states at $E_{\mathrm{F}}$ (Korringa process). However, we have to postulate correlation effects as an origin for the $T^{-1/2}$ behavior below $T\text{*}$.
For $T$ > $T\text{*}$, by taking into account temperature dependent orbital hyperfine coupling and activation-type relaxation processes to the W2 points,
we were able to explain the $T^{\mathrm{2}}$ behavior in a convincing way. 
For the unique case of the 
Ta-based WSM, we have shown that the NQR method is a direct local probe for low energy Weyl fermion excitations in the bulk. 
This is rather important because such excitations are one of the main ingredients for the unconventional electrons transport found in these new materials.  
It would be interesting to take a deeper look into other conventional bulk probes such as thermo-power to explore signatures of Weyl fermions, 
and to extend the NQR study to other Ta-based Weyl- and Dirac- semi metals.

$Acknowledgement$ We thank B. D{\'o}ra for discussions and critical reading of the manuscript. Discussions with H. Harima, K. Kanoda, M. Majumder 
and E. Hassinger are also appreciated. T. K., Y. K. and H.T. appreciate the financial support from JSPS KAKENHI Grants (Nos.15K21732 and 15H05885).


\begin{thebibliography}{30}%
\makeatletter
\providecommand \@ifxundefined [1]{%
 \@ifx{#1\undefined}
}%
\providecommand \@ifnum [1]{%
 \ifnum #1\expandafter \@firstoftwo
 \else \expandafter \@secondoftwo
 \fi
}%
\providecommand \@ifx [1]{%
 \ifx #1\expandafter \@firstoftwo
 \else \expandafter \@secondoftwo
 \fi
}%
\providecommand \natexlab [1]{#1}%
\providecommand \enquote  [1]{``#1''}%
\providecommand \bibnamefont  [1]{#1}%
\providecommand \bibfnamefont [1]{#1}%
\providecommand \citenamefont [1]{#1}%
\providecommand \href@noop [0]{\@secondoftwo}%
\providecommand \href [0]{\begingroup \@sanitize@url \@href}%
\providecommand \@href[1]{\@@startlink{#1}\@@href}%
\providecommand \@@href[1]{\endgroup#1\@@endlink}%
\providecommand \@sanitize@url [0]{\catcode `\\12\catcode `\$12\catcode
  `\&12\catcode `\#12\catcode `\^12\catcode `\_12\catcode `\%12\relax}%
\providecommand \@@startlink[1]{}%
\providecommand \@@endlink[0]{}%
\providecommand \url  [0]{\begingroup\@sanitize@url \@url }%
\providecommand \@url [1]{\endgroup\@href {#1}{\urlprefix }}%
\providecommand \urlprefix  [0]{URL }%
\providecommand \Eprint [0]{\href }%
\providecommand \doibase [0]{http://dx.doi.org/}%
\providecommand \selectlanguage [0]{\@gobble}%
\providecommand \bibinfo  [0]{\@secondoftwo}%
\providecommand \bibfield  [0]{\@secondoftwo}%
\providecommand \translation [1]{[#1]}%
\providecommand \BibitemOpen [0]{}%
\providecommand \bibitemStop [0]{}%
\providecommand \bibitemNoStop [0]{.\EOS\space}%
\providecommand \EOS [0]{\spacefactor3000\relax}%
\providecommand \BibitemShut  [1]{\csname bibitem#1\endcsname}%
\let\auto@bib@innerbib\@empty
\bibitem [{\citenamefont {Castro~Neto}\ \emph {et~al.}(2009)\citenamefont
  {Castro~Neto}, \citenamefont {Guinea}, \citenamefont {Peres}, \citenamefont
  {Novoselov},\ and\ \citenamefont {Geim}}]{Castro_2009}%
  \BibitemOpen
  \bibfield  {author} {\bibinfo {author} {\bibfnamefont {A.~H.}\ \bibnamefont
  {Castro~Neto}}, \bibinfo {author} {\bibfnamefont {F.}~\bibnamefont {Guinea}},
  \bibinfo {author} {\bibfnamefont {N.~M.~R.}\ \bibnamefont {Peres}}, \bibinfo
  {author} {\bibfnamefont {K.}~\bibnamefont {Novoselov}}, \ and\ \bibinfo
  {author} {\bibfnamefont {A.}~\bibnamefont {Geim}},\ }\href@noop {} {\bibfield
   {journal} {\bibinfo  {journal} {Rev. Mod. Phys.}\ }\textbf {\bibinfo
  {volume} {81}},\ \bibinfo {pages} {109} (\bibinfo {year} {2009})}\BibitemShut
  {NoStop}%
\bibitem [{\citenamefont {Kane}\ and\ \citenamefont {Mele}(2005)}]{Kane_2005}%
  \BibitemOpen
  \bibfield  {author} {\bibinfo {author} {\bibfnamefont {C.~L.}\ \bibnamefont
  {Kane}}\ and\ \bibinfo {author} {\bibfnamefont {E.~J.}\ \bibnamefont
  {Mele}},\ }\href@noop {} {\bibfield  {journal} {\bibinfo  {journal} {Phys.
  Rev. Lett.}\ }\textbf {\bibinfo {volume} {95}},\ \bibinfo {pages} {146802}
  (\bibinfo {year} {2005})}\BibitemShut {NoStop}%
\bibitem [{\citenamefont {Bernevig}\ \emph {et~al.}(2006)\citenamefont
  {Bernevig}, \citenamefont {Hughes},\ and\ \citenamefont
  {Zhang}}]{Bernevig_2006}%
  \BibitemOpen
  \bibfield  {author} {\bibinfo {author} {\bibfnamefont {B.~A.}\ \bibnamefont
  {Bernevig}}, \bibinfo {author} {\bibfnamefont {T.~L.}\ \bibnamefont
  {Hughes}}, \ and\ \bibinfo {author} {\bibfnamefont {S.~C.}\ \bibnamefont
  {Zhang}},\ }\href@noop {} {\bibfield  {journal} {\bibinfo  {journal}
  {Science}\ }\textbf {\bibinfo {volume} {314}},\ \bibinfo {pages} {1757}
  (\bibinfo {year} {2006})}\BibitemShut {NoStop}%
\bibitem [{\citenamefont {Hasan}\ and\ \citenamefont
  {Kane}(2010)}]{Hasan_2010}%
  \BibitemOpen
  \bibfield  {author} {\bibinfo {author} {\bibfnamefont {M.~Z.}\ \bibnamefont
  {Hasan}}\ and\ \bibinfo {author} {\bibfnamefont {C.~L.}\ \bibnamefont
  {Kane}},\ }\href@noop {} {\bibfield  {journal} {\bibinfo  {journal} {Rev.
  Mod. Phys.}\ }\textbf {\bibinfo {volume} {82}},\ \bibinfo {pages} {3405}
  (\bibinfo {year} {2010})}\BibitemShut {NoStop}%
\bibitem [{\citenamefont {Wehling}\ \emph {et~al.}(2014)\citenamefont
  {Wehling}, \citenamefont {Black-Schaffer},\ and\ \citenamefont
  {Balatsky}}]{Wehling_2014}%
  \BibitemOpen
  \bibfield  {author} {\bibinfo {author} {\bibfnamefont {T.~O.}\ \bibnamefont
  {Wehling}}, \bibinfo {author} {\bibfnamefont {A.~M.}\ \bibnamefont
  {Black-Schaffer}}, \ and\ \bibinfo {author} {\bibfnamefont {A.~V.}\
  \bibnamefont {Balatsky}},\ }\href@noop {} {\bibfield  {journal} {\bibinfo
  {journal} {Advances in Phys.}\ }\textbf {\bibinfo {volume} {63}},\ \bibinfo
  {pages} {1} (\bibinfo {year} {2014})}\BibitemShut {NoStop}%
\bibitem [{\citenamefont {Young}\ \emph {et~al.}(2012)\citenamefont {Young},
  \citenamefont {Zaheer}, \citenamefont {Teo}, \citenamefont {Kane},
  \citenamefont {Mele},\ and\ \citenamefont {Rappe}}]{Young_2012}%
  \BibitemOpen
  \bibfield  {author} {\bibinfo {author} {\bibfnamefont {S.~M.}\ \bibnamefont
  {Young}}, \bibinfo {author} {\bibfnamefont {S.}~\bibnamefont {Zaheer}},
  \bibinfo {author} {\bibfnamefont {J.~C.~Y.}\ \bibnamefont {Teo}}, \bibinfo
  {author} {\bibfnamefont {C.~L.}\ \bibnamefont {Kane}}, \bibinfo {author}
  {\bibfnamefont {E.~J.}\ \bibnamefont {Mele}}, \ and\ \bibinfo {author}
  {\bibfnamefont {A.~M.}\ \bibnamefont {Rappe}},\ }\href {\doibase
  10.1103/PhysRevLett.108.140405} {\bibfield  {journal} {\bibinfo  {journal}
  {Phys. Rev. Lett.}\ }\textbf {\bibinfo {volume} {108}},\ \bibinfo {pages}
  {140405} (\bibinfo {year} {2012})}\BibitemShut {NoStop}%
\bibitem [{\citenamefont {Wang}\ \emph {et~al.}(2013)\citenamefont {Wang},
  \citenamefont {Weng}, \citenamefont {Wu}, \citenamefont {Dai},\ and\
  \citenamefont {Fang}}]{Wang_2013}%
  \BibitemOpen
  \bibfield  {author} {\bibinfo {author} {\bibfnamefont {Z.}~\bibnamefont
  {Wang}}, \bibinfo {author} {\bibfnamefont {H.}~\bibnamefont {Weng}}, \bibinfo
  {author} {\bibfnamefont {Q.}~\bibnamefont {Wu}}, \bibinfo {author}
  {\bibfnamefont {X.}~\bibnamefont {Dai}}, \ and\ \bibinfo {author}
  {\bibfnamefont {Z.}~\bibnamefont {Fang}},\ }\href {\doibase
  10.1103/PhysRevB.88.125427} {\bibfield  {journal} {\bibinfo  {journal} {Phys.
  Rev. B}\ }\textbf {\bibinfo {volume} {88}},\ \bibinfo {pages} {125427}
  (\bibinfo {year} {2013})}\BibitemShut {NoStop}%
\bibitem [{\citenamefont {Weng}\ \emph {et~al.}(2015)\citenamefont {Weng},
  \citenamefont {Fang}, \citenamefont {Fang}, \citenamefont {Bernevig},\ and\
  \citenamefont {Dai}}]{Weng_2015}%
  \BibitemOpen
  \bibfield  {author} {\bibinfo {author} {\bibfnamefont {H.}~\bibnamefont
  {Weng}}, \bibinfo {author} {\bibfnamefont {C.}~\bibnamefont {Fang}}, \bibinfo
  {author} {\bibfnamefont {Z.}~\bibnamefont {Fang}}, \bibinfo {author}
  {\bibfnamefont {B.~A.}\ \bibnamefont {Bernevig}}, \ and\ \bibinfo {author}
  {\bibfnamefont {X.}~\bibnamefont {Dai}},\ }\href {\doibase
  10.1103/PhysRevX.5.011029} {\bibfield  {journal} {\bibinfo  {journal} {Phys.
  Rev. X}\ }\textbf {\bibinfo {volume} {5}},\ \bibinfo {pages} {011029}
  (\bibinfo {year} {2015})}\BibitemShut {NoStop}%
\bibitem [{\citenamefont {Shekhar}\ \emph {et~al.}(2015)\citenamefont
  {Shekhar}, \citenamefont {Nayak}, \citenamefont {Sun}, \citenamefont
  {Schmidt}, \citenamefont {Nicklas}, \citenamefont {Leermakers}, \citenamefont
  {Zeitler}, \citenamefont {Skourski}, \citenamefont {Wosnitza}, \citenamefont
  {Liu}, \citenamefont {Chen}, \citenamefont {Schnelle}, \citenamefont
  {Borrmann}, \citenamefont {Grin}, \citenamefont {Felser},\ and\ \citenamefont
  {Yan}}]{Shekar_2015}%
  \BibitemOpen
  \bibfield  {author} {\bibinfo {author} {\bibfnamefont {C.}~\bibnamefont
  {Shekhar}}, \bibinfo {author} {\bibfnamefont {A.~K.}\ \bibnamefont {Nayak}},
  \bibinfo {author} {\bibfnamefont {Y.}~\bibnamefont {Sun}}, \bibinfo {author}
  {\bibfnamefont {M.}~\bibnamefont {Schmidt}}, \bibinfo {author} {\bibfnamefont
  {M.}~\bibnamefont {Nicklas}}, \bibinfo {author} {\bibfnamefont
  {I.}~\bibnamefont {Leermakers}}, \bibinfo {author} {\bibfnamefont
  {U.}~\bibnamefont {Zeitler}}, \bibinfo {author} {\bibfnamefont
  {Y.}~\bibnamefont {Skourski}}, \bibinfo {author} {\bibfnamefont
  {J.}~\bibnamefont {Wosnitza}}, \bibinfo {author} {\bibfnamefont
  {Z.}~\bibnamefont {Liu}}, \bibinfo {author} {\bibfnamefont {Y.}~\bibnamefont
  {Chen}}, \bibinfo {author} {\bibfnamefont {W.}~\bibnamefont {Schnelle}},
  \bibinfo {author} {\bibfnamefont {H.}~\bibnamefont {Borrmann}}, \bibinfo
  {author} {\bibfnamefont {Y.}~\bibnamefont {Grin}}, \bibinfo {author}
  {\bibfnamefont {C.}~\bibnamefont {Felser}}, \ and\ \bibinfo {author}
  {\bibfnamefont {B.}~\bibnamefont {Yan}},\ }\href {\doibase
  {10.1038/NPHYS3372}} {\bibfield  {journal} {\bibinfo  {journal} {{Nature
  Physics}}\ }\textbf {\bibinfo {volume} {{11}}},\ \bibinfo {pages} {{645+}}
  (\bibinfo {year} {{2015}})}\BibitemShut {NoStop}%
\bibitem [{\citenamefont {Sun}\ \emph {et~al.}(2015)\citenamefont {Sun},
  \citenamefont {Wu},\ and\ \citenamefont {Yan}}]{Sun_2015}%
  \BibitemOpen
  \bibfield  {author} {\bibinfo {author} {\bibfnamefont {Y.}~\bibnamefont
  {Sun}}, \bibinfo {author} {\bibfnamefont {S.-C.}\ \bibnamefont {Wu}}, \ and\
  \bibinfo {author} {\bibfnamefont {B.}~\bibnamefont {Yan}},\ }\href {\doibase
  10.1103/PhysRevB.92.115428} {\bibfield  {journal} {\bibinfo  {journal} {Phys.
  Rev. B}\ }\textbf {\bibinfo {volume} {92}},\ \bibinfo {pages} {115428}
  (\bibinfo {year} {2015})}\BibitemShut {NoStop}%
\bibitem [{\citenamefont {Xu}\ \emph {et~al.}(2015)\citenamefont {Xu},
  \citenamefont {Belopolski}, \citenamefont {Sanchez}, \citenamefont {Zhang},
  \citenamefont {Chang}, \citenamefont {Guo}, \citenamefont {Bian},
  \citenamefont {Yuan}, \citenamefont {Lu}, \citenamefont {Chang},
  \citenamefont {Shibayev}, \citenamefont {Prokopovych}, \citenamefont
  {Alidoust}, \citenamefont {Zheng}, \citenamefont {Lee}, \citenamefont
  {Huang}, \citenamefont {Sankar}, \citenamefont {Chou}, \citenamefont {Hsu},
  \citenamefont {Jeng}, \citenamefont {Bansil}, \citenamefont {Neupert},
  \citenamefont {Strocov}, \citenamefont {Lin}, \citenamefont {Jia},\ and\
  \citenamefont {Hasan}}]{Xu_2015}%
  \BibitemOpen
  \bibfield  {author} {\bibinfo {author} {\bibfnamefont {S.-Y.}\ \bibnamefont
  {Xu}}, \bibinfo {author} {\bibfnamefont {I.}~\bibnamefont {Belopolski}},
  \bibinfo {author} {\bibfnamefont {D.~S.}\ \bibnamefont {Sanchez}}, \bibinfo
  {author} {\bibfnamefont {C.}~\bibnamefont {Zhang}}, \bibinfo {author}
  {\bibfnamefont {G.}~\bibnamefont {Chang}}, \bibinfo {author} {\bibfnamefont
  {C.}~\bibnamefont {Guo}}, \bibinfo {author} {\bibfnamefont {G.}~\bibnamefont
  {Bian}}, \bibinfo {author} {\bibfnamefont {Z.}~\bibnamefont {Yuan}}, \bibinfo
  {author} {\bibfnamefont {H.}~\bibnamefont {Lu}}, \bibinfo {author}
  {\bibfnamefont {T.-R.}\ \bibnamefont {Chang}}, \bibinfo {author}
  {\bibfnamefont {P.~P.}\ \bibnamefont {Shibayev}}, \bibinfo {author}
  {\bibfnamefont {M.~L.}\ \bibnamefont {Prokopovych}}, \bibinfo {author}
  {\bibfnamefont {N.}~\bibnamefont {Alidoust}}, \bibinfo {author}
  {\bibfnamefont {H.}~\bibnamefont {Zheng}}, \bibinfo {author} {\bibfnamefont
  {C.-C.}\ \bibnamefont {Lee}}, \bibinfo {author} {\bibfnamefont {S.-M.}\
  \bibnamefont {Huang}}, \bibinfo {author} {\bibfnamefont {R.}~\bibnamefont
  {Sankar}}, \bibinfo {author} {\bibfnamefont {F.}~\bibnamefont {Chou}},
  \bibinfo {author} {\bibfnamefont {C.-H.}\ \bibnamefont {Hsu}}, \bibinfo
  {author} {\bibfnamefont {H.-T.}\ \bibnamefont {Jeng}}, \bibinfo {author}
  {\bibfnamefont {A.}~\bibnamefont {Bansil}}, \bibinfo {author} {\bibfnamefont
  {T.}~\bibnamefont {Neupert}}, \bibinfo {author} {\bibfnamefont {V.~N.}\
  \bibnamefont {Strocov}}, \bibinfo {author} {\bibfnamefont {H.}~\bibnamefont
  {Lin}}, \bibinfo {author} {\bibfnamefont {S.}~\bibnamefont {Jia}}, \ and\
  \bibinfo {author} {\bibfnamefont {M.~Z.}\ \bibnamefont {Hasan}},\ }\href
  {\doibase 10.1126/sciadv.1501092} {\bibfield  {journal} {\bibinfo  {journal}
  {Science Advances}\ }\textbf {\bibinfo {volume} {1}} (\bibinfo {year}
  {2015}),\ 10.1126/sciadv.1501092}\BibitemShut {NoStop}%
\bibitem [{\citenamefont {Arnold}\ \emph {et~al.}(2016)\citenamefont {Arnold},
  \citenamefont {Shekhar}, \citenamefont {Wu}, \citenamefont {Sun},
  \citenamefont {dos Reis}, \citenamefont {Kumar}, \citenamefont {Naumann},
  \citenamefont {Ajeesh}, \citenamefont {Schmidt}, \citenamefont {Grushin},
  \citenamefont {Bardarson}, \citenamefont {Baenitz}, \citenamefont {Sokolov},
  \citenamefont {Borrmann}, \citenamefont {Nicklas}, \citenamefont {Felser},
  \citenamefont {Hassinger},\ and\ \citenamefont {Yan}}]{Arnold_2016}%
  \BibitemOpen
  \bibfield  {author} {\bibinfo {author} {\bibfnamefont {F.}~\bibnamefont
  {Arnold}}, \bibinfo {author} {\bibfnamefont {C.}~\bibnamefont {Shekhar}},
  \bibinfo {author} {\bibfnamefont {S.-C.}\ \bibnamefont {Wu}}, \bibinfo
  {author} {\bibfnamefont {Y.}~\bibnamefont {Sun}}, \bibinfo {author}
  {\bibfnamefont {R.~D.}\ \bibnamefont {dos Reis}}, \bibinfo {author}
  {\bibfnamefont {N.}~\bibnamefont {Kumar}}, \bibinfo {author} {\bibfnamefont
  {M.}~\bibnamefont {Naumann}}, \bibinfo {author} {\bibfnamefont {M.~O.}\
  \bibnamefont {Ajeesh}}, \bibinfo {author} {\bibfnamefont {M.}~\bibnamefont
  {Schmidt}}, \bibinfo {author} {\bibfnamefont {A.~G.}\ \bibnamefont
  {Grushin}}, \bibinfo {author} {\bibfnamefont {J.~H.}\ \bibnamefont
  {Bardarson}}, \bibinfo {author} {\bibfnamefont {M.}~\bibnamefont {Baenitz}},
  \bibinfo {author} {\bibfnamefont {D.}~\bibnamefont {Sokolov}}, \bibinfo
  {author} {\bibfnamefont {H.}~\bibnamefont {Borrmann}}, \bibinfo {author}
  {\bibfnamefont {M.}~\bibnamefont {Nicklas}}, \bibinfo {author} {\bibfnamefont
  {C.}~\bibnamefont {Felser}}, \bibinfo {author} {\bibfnamefont
  {E.}~\bibnamefont {Hassinger}}, \ and\ \bibinfo {author} {\bibfnamefont
  {B.}~\bibnamefont {Yan}},\ }\href {\doibase {10.1038/ncomms11615}} {\bibfield
   {journal} {\bibinfo  {journal} {{Nature Comm.}}\ }\textbf {\bibinfo {volume}
  {{7}}} (\bibinfo {year} {{2016}}),\ {10.1038/ncomms11615}}\BibitemShut
  {NoStop}%
\bibitem [{\citenamefont {Curro}(2011)}]{Curro_2011}%
  \BibitemOpen
  \bibfield  {author} {\bibinfo {author} {\bibfnamefont {N.}~\bibnamefont
  {Curro}},\ }\href@noop {} {\emph {\bibinfo {title} {"Quadruplole NMR of
  superconductors" in Encyclopedia of Magnetic Resonance}}}\ (\bibinfo
  {publisher} {John Wiley \& Sons, Ltd.},\ \bibinfo {year} {2011})\BibitemShut
  {NoStop}%
\bibitem [{\citenamefont {Curro}(2009)}]{Curro_2009}%
  \BibitemOpen
  \bibfield  {author} {\bibinfo {author} {\bibfnamefont {N.~J.}\ \bibnamefont
  {Curro}},\ }\href {http://stacks.iop.org/0034-4885/72/i=2/a=026502}
  {\bibfield  {journal} {\bibinfo  {journal} {Reports on Progress in Physics}\
  }\textbf {\bibinfo {volume} {72}},\ \bibinfo {pages} {026502} (\bibinfo
  {year} {2009})}\BibitemShut {NoStop}%
\bibitem [{\citenamefont {Walstedt}(2008)}]{Walstedt_2008}%
  \BibitemOpen
  \bibfield  {author} {\bibinfo {author} {\bibfnamefont {R.~E.}\ \bibnamefont
  {Walstedt}},\ }\href@noop {} {\emph {\bibinfo {title} {The NMR Probe of
  High-T$_c$ Materials}}},\ \bibinfo {series} {Springer Tracts in Modern
  Physics}, Vol.\ \bibinfo {volume} {228}\ (\bibinfo  {publisher}
  {Springer-Verlag Berlin Heidelberg},\ \bibinfo {year} {2008})\BibitemShut
  {NoStop}%
\bibitem [{\citenamefont {Kuramoto}\ and\ \citenamefont
  {Kitaoka}(2000)}]{Kuramoto_2000}%
  \BibitemOpen
  \bibfield  {author} {\bibinfo {author} {\bibfnamefont {Y.}~\bibnamefont
  {Kuramoto}}\ and\ \bibinfo {author} {\bibfnamefont {Y.}~\bibnamefont
  {Kitaoka}},\ }\href@noop {} {\emph {\bibinfo {title} {Dynamics of heavy
  electrons}}}\ (\bibinfo  {publisher} {Oxford University Press INC. New
  York},\ \bibinfo {year} {2000})\BibitemShut {NoStop}%
\bibitem [{\citenamefont {Aeppli}\ and\ \citenamefont
  {Fisk}(1992)}]{Aeppli_1992}%
  \BibitemOpen
  \bibfield  {author} {\bibinfo {author} {\bibfnamefont {G.}~\bibnamefont
  {Aeppli}}\ and\ \bibinfo {author} {\bibfnamefont {Z.}~\bibnamefont {Fisk}},\
  }\href@noop {} {\bibfield  {journal} {\bibinfo  {journal} {Comments Condens.
  Matter Phys.}\ }\textbf {\bibinfo {volume} {16}},\ \bibinfo {pages} {155}
  (\bibinfo {year} {1992})}\BibitemShut {NoStop}%
\bibitem [{\citenamefont {Riseborough}(2000)}]{Riseborough_2000}%
  \BibitemOpen
  \bibfield  {author} {\bibinfo {author} {\bibfnamefont {P.}~\bibnamefont
  {Riseborough}},\ }\href@noop {} {\bibfield  {journal} {\bibinfo  {journal}
  {Adv. Phys.}\ }\textbf {\bibinfo {volume} {49}},\ \bibinfo {pages} {257}
  (\bibinfo {year} {2000})}\BibitemShut {NoStop}%
\bibitem [{\citenamefont {Caldwell}\ \emph {et~al.}(2007)\citenamefont
  {Caldwell}, \citenamefont {Reyes}, \citenamefont {Moulton}, \citenamefont
  {Kuhns}, \citenamefont {Hoch}, \citenamefont {Schlottmann},\ and\
  \citenamefont {Fisk}}]{Caldwell_2007}%
  \BibitemOpen
  \bibfield  {author} {\bibinfo {author} {\bibfnamefont {T.}~\bibnamefont
  {Caldwell}}, \bibinfo {author} {\bibfnamefont {A.~P.}\ \bibnamefont {Reyes}},
  \bibinfo {author} {\bibfnamefont {W.~G.}\ \bibnamefont {Moulton}}, \bibinfo
  {author} {\bibfnamefont {P.~L.}\ \bibnamefont {Kuhns}}, \bibinfo {author}
  {\bibfnamefont {M.~J.~R.}\ \bibnamefont {Hoch}}, \bibinfo {author}
  {\bibfnamefont {P.}~\bibnamefont {Schlottmann}}, \ and\ \bibinfo {author}
  {\bibfnamefont {Z.}~\bibnamefont {Fisk}},\ }\href {\doibase
  10.1103/PhysRevB.75.075106} {\bibfield  {journal} {\bibinfo  {journal} {Phys.
  Rev. B}\ }\textbf {\bibinfo {volume} {75}},\ \bibinfo {pages} {075106}
  (\bibinfo {year} {2007})}\BibitemShut {NoStop}%
\bibitem [{\citenamefont {Br\"uning}\ \emph {et~al.}(2010)\citenamefont
  {Br\"uning}, \citenamefont {Brando}, \citenamefont {Baenitz}, \citenamefont
  {Bentien}, \citenamefont {Strydom}, \citenamefont {Walstedt},\ and\
  \citenamefont {Steglich}}]{Bruning_2010}%
  \BibitemOpen
  \bibfield  {author} {\bibinfo {author} {\bibfnamefont {E.~M.}\ \bibnamefont
  {Br\"uning}}, \bibinfo {author} {\bibfnamefont {M.}~\bibnamefont {Brando}},
  \bibinfo {author} {\bibfnamefont {M.}~\bibnamefont {Baenitz}}, \bibinfo
  {author} {\bibfnamefont {A.}~\bibnamefont {Bentien}}, \bibinfo {author}
  {\bibfnamefont {A.~M.}\ \bibnamefont {Strydom}}, \bibinfo {author}
  {\bibfnamefont {R.~E.}\ \bibnamefont {Walstedt}}, \ and\ \bibinfo {author}
  {\bibfnamefont {F.}~\bibnamefont {Steglich}},\ }\href {\doibase
  10.1103/PhysRevB.82.125115} {\bibfield  {journal} {\bibinfo  {journal} {Phys.
  Rev. B}\ }\textbf {\bibinfo {volume} {82}},\ \bibinfo {pages} {125115}
  (\bibinfo {year} {2010})}\BibitemShut {NoStop}%
\bibitem [{\citenamefont {{Okv{\'a}tovity}}\ \emph {et~al.}(2016)\citenamefont
  {{Okv{\'a}tovity}}, \citenamefont {{Simon}},\ and\ \citenamefont
  {{D{\'o}ra}}}]{Okvatovity_2016}%
  \BibitemOpen
  \bibfield  {author} {\bibinfo {author} {\bibfnamefont {Z.}~\bibnamefont
  {{Okv{\'a}tovity}}}, \bibinfo {author} {\bibfnamefont {F.}~\bibnamefont
  {{Simon}}}, \ and\ \bibinfo {author} {\bibfnamefont {B.}~\bibnamefont
  {{D{\'o}ra}}},\ }\href@noop {} {\bibfield  {journal} {\bibinfo  {journal}
  {ArXiv e-prints}\ } (\bibinfo {year} {2016})},\ \Eprint
  {http://arxiv.org/abs/1609.03370} {arXiv:1609.03370 [cond-mat.mes-hall]}
  \BibitemShut {NoStop}%
\bibitem [{\citenamefont {Mitrovi\ifmmode~\acute{c}\else \'{c}\fi{}}\ \emph
  {et~al.}(2001)\citenamefont {Mitrovi\ifmmode~\acute{c}\else \'{c}\fi{}},
  \citenamefont {Sigmund},\ and\ \citenamefont {Halperin}}]{Mit_2001}%
  \BibitemOpen
  \bibfield  {author} {\bibinfo {author} {\bibfnamefont {V.~F.}\ \bibnamefont
  {Mitrovi\ifmmode~\acute{c}\else \'{c}\fi{}}}, \bibinfo {author}
  {\bibfnamefont {E.~E.}\ \bibnamefont {Sigmund}}, \ and\ \bibinfo {author}
  {\bibfnamefont {W.~P.}\ \bibnamefont {Halperin}},\ }\href {\doibase
  10.1103/PhysRevB.64.024520} {\bibfield  {journal} {\bibinfo  {journal} {Phys.
  Rev. B}\ }\textbf {\bibinfo {volume} {64}},\ \bibinfo {pages} {024520}
  (\bibinfo {year} {2001})}\BibitemShut {NoStop}%
\bibitem [{\citenamefont {Chepin}\ and\ \citenamefont
  {Ross}(1991)}]{Chepin_1991}%
  \BibitemOpen
  \bibfield  {author} {\bibinfo {author} {\bibfnamefont {J.}~\bibnamefont
  {Chepin}}\ and\ \bibinfo {author} {\bibfnamefont {J.~H.}\ \bibnamefont
  {Ross}},\ }\href@noop {} {\bibfield  {journal} {\bibinfo  {journal} {J.
  Phys.: Condens. Matter}\ }\textbf {\bibinfo {volume} {3}},\ \bibinfo {pages}
  {8103} (\bibinfo {year} {1991})}\BibitemShut {NoStop}%
\bibitem [{\citenamefont {Abragam}(1961)}]{Abragam_1961}%
  \BibitemOpen
  \bibfield  {author} {\bibinfo {author} {\bibfnamefont {A.}~\bibnamefont
  {Abragam}},\ }\href@noop {} {\emph {\bibinfo {title} {The Principle of
  Nuclear Magnetism}}}\ (\bibinfo  {publisher} {Cladendon, Oxford},\ \bibinfo
  {year} {1961})\BibitemShut {NoStop}%
\bibitem [{\citenamefont {Slichter}(1989)}]{Slichter_1989}%
  \BibitemOpen
  \bibfield  {author} {\bibinfo {author} {\bibfnamefont {C.~P.}\ \bibnamefont
  {Slichter}},\ }\href@noop {} {\emph {\bibinfo {title} {Principles of Magnetic
  Resonance}}}\ (\bibinfo  {publisher} {Springer, New York},\ \bibinfo {year}
  {1989})\BibitemShut {NoStop}%
\bibitem [{\citenamefont {Koepernik}\ and\ \citenamefont
  {Eschrig}(1999)}]{Koepernik_1999}%
  \BibitemOpen
  \bibfield  {author} {\bibinfo {author} {\bibfnamefont {K.}~\bibnamefont
  {Koepernik}}\ and\ \bibinfo {author} {\bibfnamefont {H.}~\bibnamefont
  {Eschrig}},\ }\href {\doibase 10.1103/PhysRevB.59.1743} {\bibfield  {journal}
  {\bibinfo  {journal} {Phys. Rev. B}\ }\textbf {\bibinfo {volume} {59}},\
  \bibinfo {pages} {1743} (\bibinfo {year} {1999})}\BibitemShut {NoStop}%
\bibitem [{\citenamefont {Perdew}\ and\ \citenamefont
  {Wang}(1992)}]{Perdew_1992}%
  \BibitemOpen
  \bibfield  {author} {\bibinfo {author} {\bibfnamefont {J.~P.}\ \bibnamefont
  {Perdew}}\ and\ \bibinfo {author} {\bibfnamefont {Y.}~\bibnamefont {Wang}},\
  }\href {\doibase 10.1103/PhysRevB.45.13244} {\bibfield  {journal} {\bibinfo
  {journal} {Phys. Rev. B}\ }\textbf {\bibinfo {volume} {45}},\ \bibinfo
  {pages} {13244} (\bibinfo {year} {1992})}\BibitemShut {NoStop}%
\bibitem [{\citenamefont {Koch}\ \emph {et~al.}(2010)\citenamefont {Koch},
  \citenamefont {Koepernik}, \citenamefont {Neck}, \citenamefont {Rosner},\
  and\ \citenamefont {Cottenier}}]{Koch_2010}%
  \BibitemOpen
  \bibfield  {author} {\bibinfo {author} {\bibfnamefont {K.}~\bibnamefont
  {Koch}}, \bibinfo {author} {\bibfnamefont {K.}~\bibnamefont {Koepernik}},
  \bibinfo {author} {\bibfnamefont {D.~V.}\ \bibnamefont {Neck}}, \bibinfo
  {author} {\bibfnamefont {H.}~\bibnamefont {Rosner}}, \ and\ \bibinfo {author}
  {\bibfnamefont {S.}~\bibnamefont {Cottenier}},\ }\href {\doibase
  10.1103/PhysRevA.81.032507} {\bibfield  {journal} {\bibinfo  {journal} {Phys.
  Rev. A}\ }\textbf {\bibinfo {volume} {81}},\ \bibinfo {pages} {032507}
  (\bibinfo {year} {2010})}\BibitemShut {NoStop}%
\bibitem [{\citenamefont {Korringa}(1950)}]{Korringa_1950}%
  \BibitemOpen
  \bibfield  {author} {\bibinfo {author} {\bibfnamefont {J.}~\bibnamefont
  {Korringa}},\ }\href@noop {} {\bibfield  {journal} {\bibinfo  {journal}
  {Physica}\ }\textbf {\bibinfo {volume} {16}},\ \bibinfo {pages} {601}
  (\bibinfo {year} {1950})}\BibitemShut {NoStop}%
\bibitem [{\citenamefont {Katayama}\ \emph {et~al.}(2007)\citenamefont
  {Katayama}, \citenamefont {Kawasaki}, \citenamefont {Nishiyama},
  \citenamefont {Sugawara}, \citenamefont {Kikuchi}, \citenamefont {Sato},\
  and\ \citenamefont {qing Zheng}}]{Katayama_2007}%
  \BibitemOpen
  \bibfield  {author} {\bibinfo {author} {\bibfnamefont {K.}~\bibnamefont
  {Katayama}}, \bibinfo {author} {\bibfnamefont {S.}~\bibnamefont {Kawasaki}},
  \bibinfo {author} {\bibfnamefont {M.}~\bibnamefont {Nishiyama}}, \bibinfo
  {author} {\bibfnamefont {H.}~\bibnamefont {Sugawara}}, \bibinfo {author}
  {\bibfnamefont {D.}~\bibnamefont {Kikuchi}}, \bibinfo {author} {\bibfnamefont
  {H.}~\bibnamefont {Sato}}, \ and\ \bibinfo {author} {\bibfnamefont
  {G.}~\bibnamefont {qing Zheng}},\ }\href {\doibase 10.1143/JPSJ.76.023701}
  {\bibfield  {journal} {\bibinfo  {journal} {Journal of the Physical Society
  of Japan}\ }\textbf {\bibinfo {volume} {76}},\ \bibinfo {pages} {023701}
  (\bibinfo {year} {2007})}\BibitemShut {NoStop}%
\end{thebibliography}

\begin{thebibliography}{99}
\bibitem{1976} J. Christiansen, P. Heubes, R. Keitel, W. Klinger, W. Loeffler,W. Sandner, and W. Witthuhn, Z. Phys. B: Condens. Matter \textbf{24},177 (1976).
\bibitem{1991} J. Chepin and J. H. Ross, J. Phys.: Condens. Matter \textbf{3}, 8103 (1991).
\bibitem{1996} J. P. Perdew, K. Burke, and M. Ernzerhof, Phys.Rev.Lett. \textbf{77}, 3865 (1996).
\bibitem{2008} A. A. Mosto, J. R. Yates, Y. S. Lee, I. Souza, D. Vanderbilt, and N. Marzari, Comput. Phys. Commun. \textbf{178}, 685 (2008).
\bibitem{2016} Hirata Michihiro, Kyohei Ishikawa, Kazuya Miyagawa, Masafumi Tamura, Claude Berthier, Denis Basko,
Akito Kobayashi, Genki Matsuno, and Kazushi Kanoda, Nature Communications \textbf{7} (2016).
\end{thebibliography}

%

\begin{widetext}
\begin{center}
{\large\bf Supplemental Material}
\end{center}
\end{widetext}

\setcounter{equation}{0}
\setcounter{figure}{0}
\setcounter{table}{0}
\setcounter{page}{1}
\makeatletter
\renewcommand{\theequation}{S\arabic{equation}}
\renewcommand{\thefigure}{S\arabic{figure}}
\renewcommand{\thesection}{S\arabic{section}}
\renewcommand{\bibnumfmt}[1]{[S#1]}
\renewcommand{\citenumfont}[1]{S#1}

\section{Temperature dependence of $\nu_{\mathrm{Q}}$ and $\eta$}

The temperature dependence of the quadrupole frequency $\nu_{\mathrm{Q}}(T)$
 generally decreases with increasing temperature (Fig.\,S1) which is correlated to the 
lattice expansion. Owing to the fact that the origin of the EFG is
the non-spherical distribution of surrounding point charges it may be easy to
 understand the thermally activated anharmonic phonon vibrations. 
This mechanism may be applicable to the
case where onsite electric contribution to the EFG is negligibly small, 
and using the
characteristic phonon frequency $\nu_{\mathrm{L}}$, $\nu_{\mathrm{Q}}(T)$ 
is expressed as 
\begin{equation}
\nu_{Q}(T) = \nu_{Q}(0) \cdot \left( 1 - \frac{\lambda h \nu_{L}}{k_{B}}\rho(T) \right)
\end{equation}
where, $\rho(T)$ is the temperature dependent phonon distribution function
given by $\rho(T)$ = coth($h \nu_{L}/2k_{B}T$)/2 and $\lambda$ is a 
lattice structure dependent parameter. Although Eq. (S1) has been applied
successfully to insulating Cu$_{2}$O, the origin of the EFG in the present
case is due to the onsite electric contribution related
directly to the symmetry of the electronic 
wave function of the bands at the nucler site. In such a case, there is
no analytical expression for $\nu_{Q}(T)$. Therefore, we adopt an
empirical form for conventional non-cubic metals,\cite{1976}
\begin{equation}
\nu_{Q}(T) = \nu_{Q}(0) \cdot (1 - A \cdot T^{3/2}), A > 0
\end{equation}
where there exists a correlation between the magnitude of $\nu_{Q}$(0)
and the strength of the EFG's temperature variation, quantified by the 
coefficient $A$. A least-squares fit of the data above 60K to Eq. (S2) 
yields
the following values: $\nu_{Q}$ = 19.252 MHz, $A$ = 3.57$\times$10$^{-5}$ K$^{-3/2}$.

\begin{figure}[t]
\includegraphics[clip,width=\columnwidth]{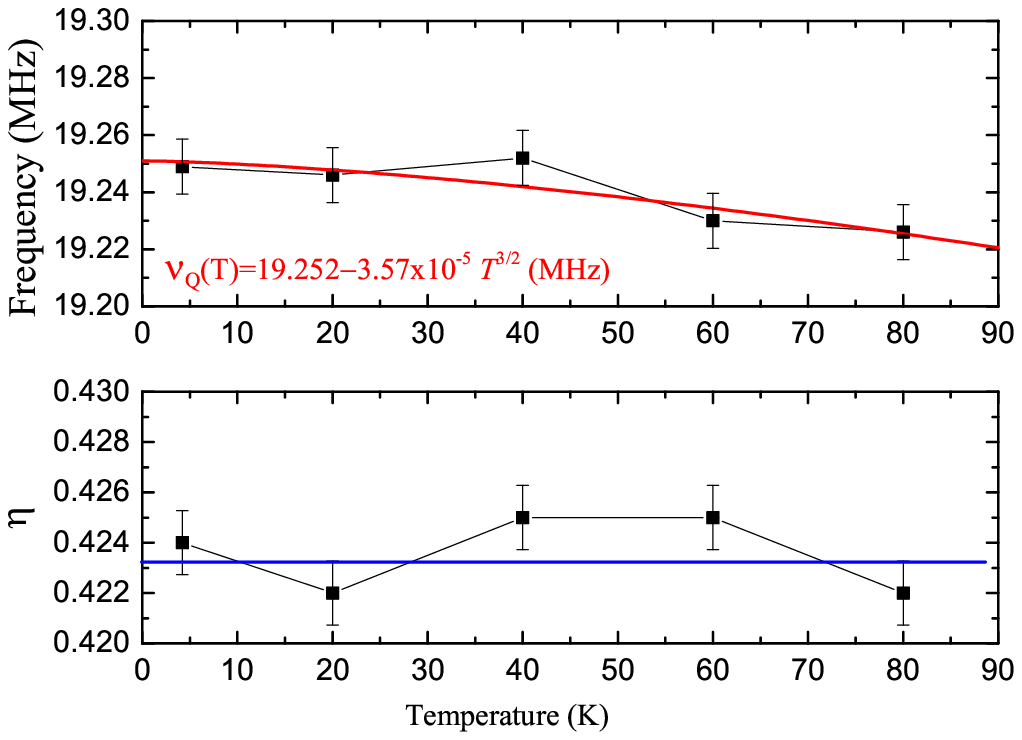}
\caption{Temperature dependence of $\nu_{Q}$ and $\eta$ in TaP.
}
\end{figure}

\section{Calculation of 1/$T_{1}T$ from non-interacting band structure 
in TaP}

We have performed $ab-initio$ density functional theory calculations
for the band structure of the TaP using the Vienna ab-initio simulation package
(VASP)\cite{1991} employing the modified Becke-Johnson (MBJ) potential\cite{1996} for the 
exchange-correlation functions. Then
we projected the DFT Bloch wave functions into the maximum localized Wannier 
functions by the
Wannier90 package.\cite{2008} Based on the tight-binding Wannier Hamiltonian, 
we interpolate the density of
states by a dense $k$-grid of 400 $\times$ 400 $\times$ 400 in the first 
Brillouin zone. Calculated $N(E)$ are shown in 
Fig. S2 (left). The nuclear spin-lattice relaxation rate for 
non-interacting electrons are calculated by the following
formula\cite{2008,2016}:
\begin{eqnarray}
\frac{1}{T_{1}} &=& \frac{\pi}{\hbar}(\gamma_{e}\gamma_{n}\hbar^{2})^{2}(A^{\perp}) \int_{-\infty}^{\infty}dED_{+}(E)f(E) \nonumber \\
                          &\cdot& \int_{-\infty}^{\infty}dE'D_{-}(E')[1 - f(E')]\delta(E - E')
\end{eqnarray}
where, $A^{\mathrm{\perp}}$ is the hyperfine coupling constant; $\gamma_{e}$, 
$\gamma_{n}$ are the electronic and nuclear gyromagnetic ratios, 
respectively. The electronic density of states for spin up and spin down
electrons is simply evaluated via Zeeman splitting $D_{\pm}$ = $E(E \mp E_{z}/2)/2$ (we take $E_{z}$ = 0 here to estimate the zero field case). $f(E)$ is
the Fermi-Dirac distribution to include temperature dependent behavior and
$\delta(E - E')$ is the delta function. 
As we are mainly interested in the temperature-dependence of $T_1$, the 
prefactors before the integral make
no difference in the general trend. Thus, we normalized arbitrarily the 
calculated values to the experimental
value at 2K, which is shown in Fig. S2 (right). 
Although the general feature is similar to the experiment, it
should be noted that the energy scale of the point where 1/$T_{1}T$ 
starts increase is roughly one order of
magnitude larger than the experimental value and the W2 energy. 
So, we can conclude that the increase of
1/$T_{1}T$
here is just due to the influence of other bands and not related to the 
excitations around Weyl points.
One of the possible
origins of this discrepancy is the lack of an adequate account of electron-electron
interactions in theory. 

\begin{figure}[t]
\includegraphics[clip,width=\columnwidth]{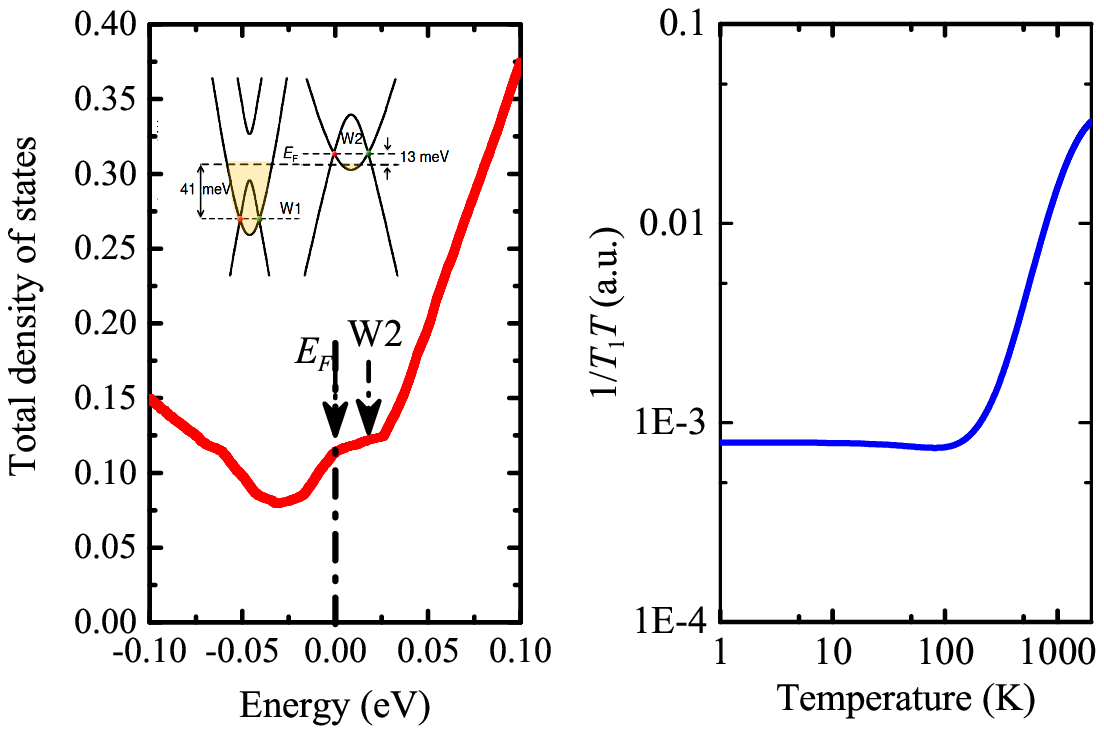}
\caption{(Left) Electronic density of states. The $E_{F}$ (=8.257 eV)
is shifted to zero. The positions of W1 and W2
Weyl points are illustrated in the inset. The W2
Weyl points are very close to $E_{F}$ (Right) The
temperature dependence of 1/$T_{1}T$ (normalized to
the experimental value at 2 K) on a logarithmic
scale. 
}
\end{figure}

\end{document}